# Beam damage of single semiconductor nanowires during X-ray nano beam diffraction experiments

Ali AlHassan[(1),*], Jonas Lähnemann[(2),*], Arman Davtyan[(1)], Mahmoud Al-Humaidi[(1)], Jesús Herranz[(2)], Danial Bahrami[(1)], Taseer Anjum[(1)], Florian Bertram[(3)], Arka Bikash Dey[(3)], Lutz Geelhaar[(2)], Ullrich Pietsch[(1)]

[(1)]Naturwissenschaftlich-Technische Fakultät der Universität Siegen, 57068 Siegen, Germany
[(2)]Paul-Drude-Institut für Festkörperelektronik, Leibniz-Institut im Forschungsverbund Berlin e.V., Hausvogteiplatz 5–7, 10117 Berlin, Germany
[(3)]DESY Photon Science, Notkestr. 85, 22607 Hamburg, Germany

(*) Both authors, Ali AlHassan and Jonas Lähnemann contributed equally to this work.

**Abstract:**

Nanoprobe X-ray diffraction (nXRD) using focused synchrotron radiation is a powerful technique to study the structural properties of individual semiconductor nanowires. However, when performing the experiment under ambient conditions, the required high X-ray dose and prolonged exposure times can lead to radiation damage. To unveil the origin of radiation damage, we compare nXRD experiments carried out on individual semiconductor nanowires in their as grown geometry both under ambient conditions and under He atmosphere at the microfocus station of the P08 beamline at the 3$^{rd}$ generation source PETRA III. Using an incident X-ray beam energy of 9 keV and photon flux of $10^{10}$ s$^{-1}$, the axial lattice parameter and tilt of individual GaAs/In$_{0.2}$Ga$_{0.8}$As/GaAs core-shell nanowires were monitored by continuously recording reciprocal space maps of the 111 Bragg reflection at a fixed spatial position over several hours. In addition, the emission properties of the (In,Ga)As quantum well, the atomic composition of the exposed nanowires and the nanowire morphology are studied by cathodoluminescence spectroscopy, energy dispersive X-ray spectroscopy and scanning electron microscopy, respectively, both prior to and after nXRD exposure. Nanowires exposed under ambient conditions show severe optical and morphological damage, which was reduced for nanowires exposed under He atmosphere. The observed damage can be largely attributed to an oxidation process from X-ray induced ozone reactions in air. Due to the lower heat transfer coefficient compared to GaAs, this oxide shell limits the heat transfer through the nanowire side facets, which is considered as the main channel of heat dissipation for nanowires in the as-grown geometry.

**Introduction:**

The development of X-ray optics and third generation synchrotron radiation sources with high brightness, nano-focused X-ray beams (Martínez-Criado *et al.* 2016, Leake *et al.* 2019) facilitates probing the structural parameters, spatial alloy distribution, crystal phases and strain distribution of single nanowires (NWs) by means of X-ray based methods. These methods include nano X-ray fluorescence (AlHassan *et al.* 2018), coherent Bragg ptychography (Hill *et al.* 2018) as well as coherent and non-coherent nanoprobe X-ray diffraction (nXRD) (Biermanns *et al.* 2013, Stankevič *et al.* 2015). For example, in AlHassan *et al.* (2018), we have demonstrated that the thicknesses of the core and shells within individual core-shell NWs, as well as the strain distribution, can be accessed in the as-grown geometry by recording reciprocal space maps (RSM) of in-plane Bragg reflections. However, during the acquisition of RSMs around Bragg reflections, the beam is fixed to a specific position on the NW for an extended time.

The continuous exposure during such experiments will inevitably increase the risk of radiation damage due to an accumulating X-ray dose at the illuminated NW section. For instance, Shi *et al.* (2012) have demonstrated radiation-induced bending of Si-on-insulator NWs by means of coherent diffraction imaging, i.e. the authors have observed a splitting of the Bragg reflections, which continuously evolves with increasing X-ray dose. Working on similar systems, Mastropietro *et al.* (2013) quantitatively demonstrated elastic strain relaxation in single Si-on-insulator lines under the influence of prolonged X-ray exposure. This was done by monitoring the evolution of the $11\bar{3}$ Bragg peak while illuminating the same position for different time intervals. The structural damage induced by the absorbed X-ray dose was shown to occur only when an oxide layer is present under the Si thin film. Whereas Polvino *et al.* (2008) describe the exposure to induce permanent structural damage to the crystal lattice, Mastropietro *et al.* (2013) have shown that the intense radiation only damages the Si/SOI interface but no the crystalline Si structure. Aside from SiC and GaN that potentially offer radiation hard alternatives to silicon devices (Sellin *et al.* 2006), bulk or layered GaAs is known to be a very radiation hard material suitable for X-ray detectors (Claeys *et al.* 2002, Lioliou *et al.* 2016, Smolyanskiy *et al.* 2018). Here, we show that GaAs/(In,Ga)As/GaAs core-shell NWs may also be affected by X-ray induced radiation damage. We will illustrate in detail the impact of the exposure to high X-ray doses on the structure, morphology and optical emission of individual NWs.

In this study, two separate experiments have been carried out using the same beam conditions, i.e. photon flux, energy and beam size, where individual NWs are continuously exposed at a fixed position along their growth axes. The first experiment is performed under air atmosphere, whereas the second is done under He atmosphere. The structural changes, e.g. tilting and axial lattice variation, were monitored by continuously recording RSMs of the 111 Bragg reflection as a function of exposure time and absorbed X-ray dose. Scanning electron microscopy (SEM) imaging is used before and after X-ray exposure to observe morphological changes, energy-dispersive X-ray spectroscopy (EDX) to track compositional changes, and hyperspectral cathodoluminesce (CL) mapping to assess the impact on the optical emission. To facilitate the latter, a NW sample with a core-shell quantum well (QW) geometry is used.

**Experimental details:**

The investigated NWs were grown by molecular beam epitaxy on a patterned Si(111) substrate using the Ga-assisted vapor-liquid-solid mechanism. The as-grown NWs are about 2.5 µm in length and 150 nm in diameter and contain radial heterostructures of GaAs/(In,Ga)As/GaAs with 20% nominal In concentration, 10 nm thickness of the (In,Ga)As QW shell, and 30 nm thickness of the outer GaAs shell. The investigated NWs were grown along a straight line on the substrate with a spacing of 10 µm between two neighboring NWs (AlHassan *et al.* 2018). This makes it possible to access the same individual NWs both in nXRD and SEM/CL/EDX measurements. More details about the growth process and sample geometry can be found in Küpers *et al.* (2018) and Küpers *et al.* (2019).

The two nXRD experiments performed under air and He atmospheres were both carried out at beamline P08 of PETRA III (Seeck *et al.* 2012) using identical conditions. A photon energy of 9 keV was used, and the photon flux integrated over the cross-section of the beam was about $10^{10}$ s$^{-1}$, while the vertical and horizontal full width at half maxima of the beam were 0.6 µm and 1.8 μm, respectively. In order to study the impact of X-ray exposure on the structural properties, NWs were measured for several hours either under air or inert gas atmosphere. A total of twelve NWs were systematically exposed for durations between 1 and 4 hours under ambient conditions. Similarly, five individual NWs were exposed for 1 to 3 hours under He atmosphere (see supporting information for details on the experimental implementation). For the four NWs discussed in this paper, the exposure times and the measurement atmospheres are listed in Table 1. To trace structural changes during the exposure, RSMs of the 111 Bragg reflection were continuously recorded at the same position along the NW growth axis. The time needed to record each RSM was about 10 minutes. The methods used to translate from real space to reciprocal space and to construct a three-dimensional (3D) RSM of the measured Bragg reflection can be found in Pietsch *et al.* (2004) and AlHassan *et al.* (2018).

*Table 1: The exposure times in hours and the atmospheres surrounding the four discussed NWs during the measurements.*

|                  | NW1 | NW2 | NW3 | NW4 |
|------------------|-----|-----|-----|-----|
| Exposure time (h) | 1   | 4   | 1   | 3   |
| Atmosphere       | Air | Air | He  | He  |

For all investigated NWs, SEM images and CL hyperspectral line-scans were recorded before and after X-ray exposure using a Zeiss Ultra55 field-emission SEM operated at an acceleration voltage of 5 kV with beam currents of 0.2–0.3 nA. For low-temperature CL measurements at 15 K, the SEM is fitted with a Gatan MonoCL4 system and a He-cooled sample stage (Lähnemann *et al.* 2019). To measure the as-grown NWs, the sample is cleaved close to the line of NWs and mounted at an angle of 45°, which is accounted for when plotting the CL. Line-scans are recorded by scanning the beam along the axis of the NW and recording a spectrum for 1 s at every dwell point. The python package *HyperSpy* is used to process the CL data (de la Peña *et al.* 2019). EDX measurements of selected unexposed and exposed NWs were recorded in the same SEM using an EDAX silicon drift detector (Apollo XV) at an acceleration voltage of 3 keV, probing the L-lines of Ga and As, as well as the K-lines of C, O and Si.

**Experimental results**

Starting with the experiment under ambient conditions (air atmosphere), the impact of X-ray exposure on the NW structure will be exemplified for two representative NWs out of twelve measured NWs. At the beginning of the exposure, the two NWs display well defined hexagonal cross-sections and side facets, as seen in Figs. 1(a),1(e).

The first NW, referred to as NW1 in Table 1, was exposed for 1 hour, which corresponds to an absorbed dose of $2.5 \times 10^{11}$ Gy (see estimate in section 3 of the supporting information). After exposure, its diameter increased by 50 nm which, if assumed symmetric, is 25 nm on the wall of each opposing side facet, and its length increased by 200 nm; see Fig. 1(b). This causes the well-defined side facets to disappear. As detailed below, we attribute this increase in the diameter and length of the NW mostly to oxidation of the GaAs initiated by the creation of ozone under the high intensity X-ray beam. Such an oxidation process under strong UV illumination in air has previously been investigated for planar GaAs layers (Flinn *et al.* 1990, Lu *et al.* 1993, Hollinger *et al.* 1994).

The morphological changes were accompanied by a local degradation in the optical properties of the NW. This degradation was visualized by spectrally-resolved CL line-scans of the emission from the (In,Ga)As QW acquired before, Fig. 1(c), and after, Fig. 1(d), exposure. Here, the emission energy is plotted against beam position along the NW axis with the emission intensity color-coded on a logarithmic color-scale. Before exposure, the NW QW showed a homogeneous distribution of the luminescence along the NW growth axis. The emission is centered at about 1.25 eV and shows only a minor blueshift between the bottom and center of the NW. The tip of the NW does not emit due to a reduced crystal quality and the absence of the QW in this segment that is formed by axial elongation during shell growth, which also leads to a change in faceting observed in Figs. 1(a),1(e) (Lähnemann *et al.* 2019). After exposure, a significant degradation of the CL emission is visible in the segment between 1.2 and 1.6 μm along the NW axis (Fig. 1(d)), which is assumed to be the position of the X-ray nanobeam.

The second NW, referred to as NW2 in Table 1, was exposed for 4 hours, corresponding to an absorbed dose of $10 \times 10^{11}$ Gy. For NW2, more severe morphological changes are observed after exposure, concerning the outer surface, length and diameter of the NW (see Figs. 1(e),1(f)). First, the NW surface lost its well defined facets, which again can be explained by ozone-induced oxidation. Second, the NW section indicated by a red dashed circle in Fig. 1(e) vanishes and beneath it, a swelling-like feature becomes visible in Fig. 1(f). As we discuss below, the top section of the NW, which is approximately 500 nm in length, has melted and formed the swelling that we observe in Fig. 1(f). As a consequence, the NW diameter is increased to approximately 210 nm at the bottom and up to 480 nm at the swollen area. The melting of the NW top section after exposure brings us to the conclusion that the NW section named P in Fig. 1(e) was illuminated by the peak of the Gaussian shaped X-ray beam, whereas the NW section circled in pink and named T was illuminated by its tail. Due to the submicron vertical size and Gaussian profile of the beam, the ozone oxidation as well as the density of the growing oxide is following the local X-ray intensity, i.e. it is maximum at the position of the center of Gaussian X-ray beam under air atmosphere and less in the tails. The morphological damage of NW2 was accompanied by a complete loss of its CL emission along the whole length of the NW (not shown).

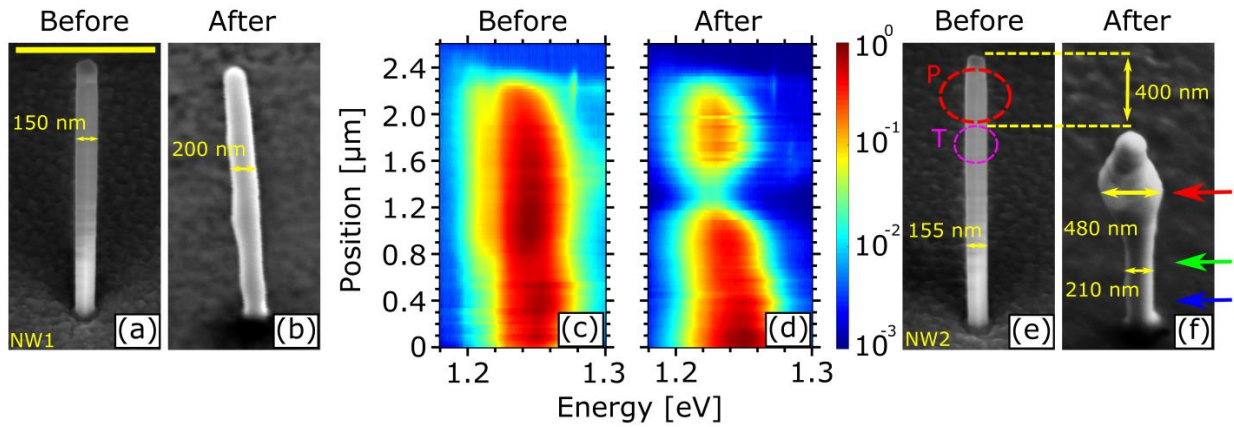

*Figure 1: (a) and (b) SEM micrographs of NW1 before and after exposure for 1 hour, respectively. (c) and (d) Normalized low-temperature hyperspectral CL line-scans along the NW growth axis before and after exposure (color-coded logarithmic intensity scale) for NW1. Note that the emission intensity depends sensitively on the positioning of the NW with respect to the focal point of the parabolic mirror so that the absolute intensities of the two line-scans cannot be compared directly. (e) and (f) SEM micrographs of NW2 before and after exposure for 4 hours. Red and pink dashed circles, indicate NW sections illuminated by the peak (P) and tail (T) of the Gaussian beam. Colored arrows indicate the positions at which the EDX measurements shown in Fig. 2 were carried out. The scale bar in (a) corresponds to 1 μm and applies to all SEM images.*

To investigate the elemental composition of the shell formed around the NWs, EDX measurements were carried out at the top, mid and bottom sections of the same NWs measured by nXRD in air. The resulting spectra for NW2 are shown in Fig. 2(a) at the positions marked by arrows in Fig. 1(f), where each spectrum was normalized with respect to the Ga peak. Compared with a NW that was not exposed to the X-ray beam, shown as horizontal pink lines in Fig. 2(a), a significant increase of the O signal is seen along the whole length of the NW. Towards the swollen top, the O signal is further enhanced and also the C peak is increased. This measurement confirms that oxidation of the NW surface takes place, which is likely to be the major driving force for the degradation of the NW morphology, and optical properties. However, it is worth noting that the sensitivity of EDX decreases for small characteristic X-ray energies and is thus lowest for the C peak. The presence of hydrocarbon molecules on the surface is inevitable. These molecules can be cracked by the impinging energetic X-ray beam during exposure leading to the deposition of amorphous C on the surface of the NW (Boller *et al.* 1983). The oxidized NW might lead to an enhanced scattering of electrons into the substrate, which would explain the slight Si signal visible for the exposed NWs.

To reduce the ozone-driven oxidation of the NW crystal structure, we replicate the nXRD experiment using the same beam size, photon flux and energy but under He atmosphere. Here, He was pumped into a cylinder made from kapton tape that has been implemented to shield the sample from air (see Fig. S1 in the supporting information). A small hole was drilled at the side wall of the kapton tape chamber to release the He overpressure. The EDX spectrum for NW4, exposed to the X-ray beam for 3 hours in the He atmosphere, is shown in Fig. 2(b). A SEM micrograph of this NW, indicating the positions of the EDX measurements, is displayed in Fig. 3(e). As expected, the O content is significantly reduced for NW4 compared to NW2. Compared with the peak heights for the non-exposed NWs indicated by the horizontal pink lines in Fig. 2, the O signal is increased only for the middle section of NW4, which is the beam position during exposure. A small O signal even under He atmosphere can be explained by residual O inside the kapton chamber.

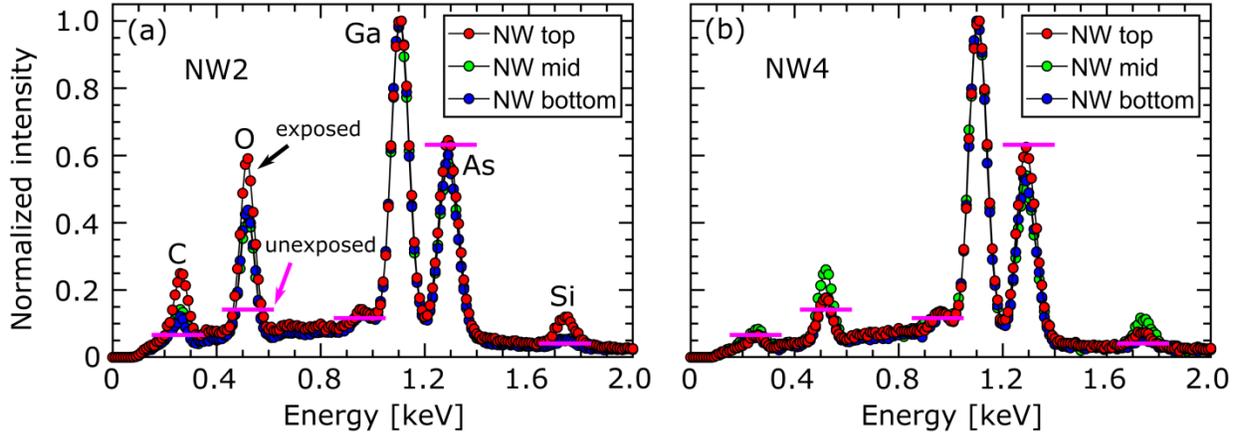

*Figure 2: EDX spectra taken at the top, mid and bottom sections of (a) NW2 (air) and (b) NW4 (He), respectively. The color of each line profile is correlated with the color of the arrows in Fig. 1(f) for NW2 and in Fig. 3(f) for NW4. The horizontal pink lines correspond to the peak height for each element in a non-exposed NW.*

To further compare the exposure under air and He atmospheres, Fig. 3 shows SEM micrographs before and after, as well as CL line-scans after X-ray exposure for two NWs measured in the He chamber: NW3, exposed for 1 hour, and NW4. In contrast to NW1, which was exposed for a similar duration in air, NW3 barely shows any increase in diameter or tilt and its facets still appear well pronounced. In addition, only a minor reduction of the CL intensity is observed after exposure at the mid-section of NW3, whereas the CL emission is locally quenched for NW1. NW4 shows a more significant shell deposition compared to NW3 (Fig. 3(b)) but much less pronounced compared to NW2 or even NW1 (Figs. 1(f),1(b)). In line with the EDX measurements, NW4 is slightly widened at its mid-section. In contrast to NWs that were exposed for more than 2 hours in air atmosphere, the CL emission of NW4 is not completely quenched, i.e. it is still visible at the bottom (Fig. 3(f)).

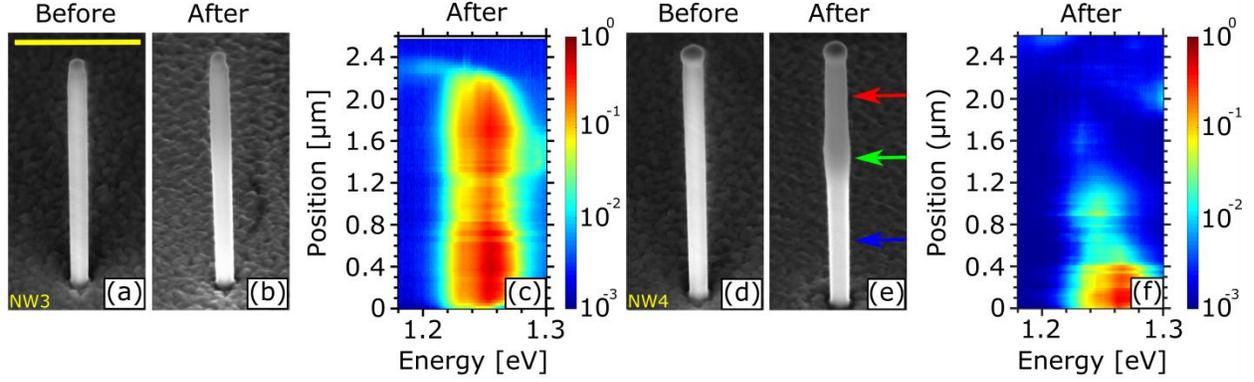

*Figure 3: (a) and (b) SEM micrographs of NW3 before and after exposure for 1 hour, respectively. (c) Normalized low-temperature hyperspectral CL line-scan along the NW growth axis after exposure (color-coded on a logarithmic intensity scale) for NW3. (e) and (f) SEM micrographs of NW4 before and after exposure for 3 hours. Colored arrows indicate the positions at which EDX measurements were carried out. (f) Normalized low-temperature hyperspectral CL line-scan after exposure for NW4. The scale bar in (a) corresponds to 1 µm and applies to all SEM images.*

## RSMs of the 111 Bragg reflection

In the following, we present a detailed analysis of nXRD measurements reflecting the beam-induced damage as function of exposure time. We continuously recorded RSMs of the 111 Bragg reflection on the individual NWs. Starting with the NWs exposed in air atmosphere, a selection of typical $(Q_Z^{111}, Q_Y^{111})$ and $(Q_Y^{111}, Q_X^{111})$ two-dimensional (2D) projections of the 3D 111 Bragg reflection of NW2 are presented in Figs. 4(a)–(e) and Figs. 4(f)–(j), respectively. Here, $Q_Z^{111}$ is defined along the scattering direction of the 111 Bragg reflection in reciprocal space and is sensitive to polytypism and variation in the axial $c$ lattice parameter. The reciprocal space vectors $Q_X^{111}$ and $Q_Y^{111}$ are defined along the $[22\bar{4}]$ and $[2\bar{2}0]$ directions of the NW, respectively, and are sensitive to both the NW thickness and tilt. The RSMs in Figs. 4(a),4(f) have been recorded only 3.6 min after the start of the exposure. The Si 111 Bragg reflection in the upper part of the RSM was considered as a reference to calculate the variation in the axial lattice spacing of the NW and therefore was placed at the unstrained position of $Q_Z^{111} = 20.038$ nm$^{-1}$. Apart from Si, the pseudomorphic GaAs ZB ($Q_Z^{111} \approx 19.22$ nm$^{-1}$) and WZ ($Q_Z^{111} \approx 19.06$ nm$^{-1}$) Bragg reflections are visible. The nominal positions for unstrained GaAs of both polytypes, $Q_Z^{111} \approx 19.25$ nm$^{-1}$ for ZB and $Q_Z^{111} \approx 19.09$ nm$^{-1}$ for WZ, are indicated by dashed Debye-Scherrer rings in all ($Q_Z^{111}$, $Q_Y^{111}$) RSMs. Both reflections are slightly shifted from the unstrained positions due to the lattice mismatch between GaAs and the (In,Ga)As shell with nominal indium content of 20%. After 48 min of X-ray exposure, the ZB reflection elongates towards smaller $Q_Z^{111}$ values and shifts along $Q_Y^{111}$ giving evidence for lattice expansion and tilt. The WZ peak also moves towards lower $Q_Z^{111}$ values. After 92 min, the main peak splits into 2 sub-peaks. Considering Fig. 1(e), the sub-peaks circled in pink and red resemble the NW sections assumed to be illuminated by the tail and peak of the Gaussian nanobeam, respectively. At the end of the exposure, the sub-peak originating from section P vanishes, which is attributed to the melting of this NW section as observed in Fig. 1(f).

In the ($Q_Y^{111}$, $Q_X^{111}$) RSMs, the small red and black circles correspond to the Si crystal truncation rod (CTR) and the NW tilt at the beginning of exposure, respectively. The thickness oscillations present in Fig. 4(f) and indicated by a green, dashed rectangle correspond to a NW diameter of 154 ± 5 nm, which is in very good agreement with the SEM observation. From the appearance of thickness oscillations of the ZB reflection, it is evident that the beam is well aligned on the NW. After 48 min, Fig. 4(g), the thickness oscillations disappear, which may be explained by the amorphous layer created by the oxidation of GaAs. Starting after 92 min of exposure, similar as for the ($Q_Z^{111}$, $Q_Y^{111}$) maps, the Bragg peak divides into two sub-peaks. The first sub-peak, denoted by T, remains at the same $Q_X^{111}$ and $Q_Y^{111}$ positions showing no tilt, whereas the second sub-peak, named P, splits from the first one and moves along the dashed arrow in Fig. 4(i). At the end of the exposure (after 240 min), P vanishes. The RSMs of NW1 are given in the supporting information (Fig. S2) and show a behavior similar to the RSMs of NW2 recorded during the first hour of exposure.

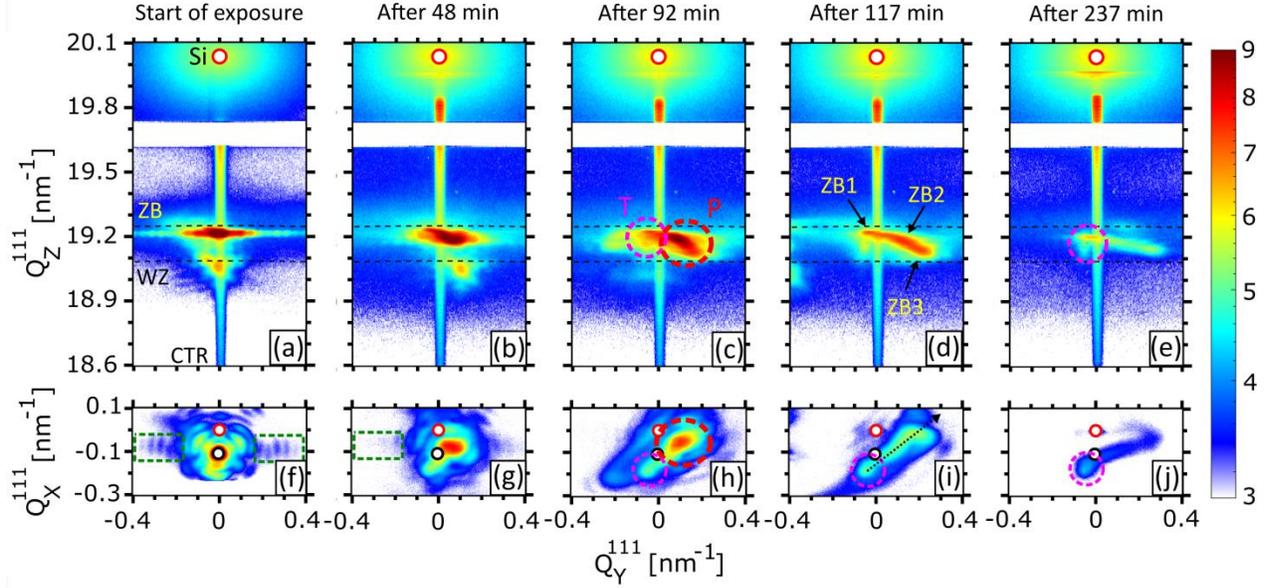

*Figure 4: (a)–(e) 2D projections of the 111 Bragg reflection in the $(Q_Z^{111}, Q_Y^{111})$ reciprocal space plane for NW2 exposed in air. The two dashed curves in $Q_Z^{111}$ represent the ZB (top) and WZ (bottom) Debye-Scherrer rings. The red and pink dotted circles named P and T are explained in Fig. 1(e) and represent sub-Bragg peaks that originate from NW sections illuminated by the peak and tail of the Gaussian beam. The peaks named ZB1–ZB3 will be explained in relation with Fig. 6(d). (f)–(j) 2D projections of the 111 Bragg reflection in the $(Q_Y^{111}, Q_X^{111})$ reciprocal space plane. The red and black circles represent the positions of the Si CTR and the initial NW tilt, respectively, in $Q_X^{111}$ and $Q_Y^{111}$. The time at which each RSM acquisition was started is mentioned at the top.*

Similarly, several individual NWs have been exposed under He atmosphere at a fixed position along their growth axes for different time intervals while continuously recording RSMs of the 111 Bragg reflection. Fig. 5 shows the $(Q_Z^{111}, Q_Y^{111})$ and $(Q_Z^{111}, Q_X^{111})$ projections of the 3D 111 Bragg reflection of NW4 that has been exposed for 3 hours. In contrast to NW2, the respective Bragg reflection stays constant in intensity and shows almost no variation along all reciprocal space directions defined by $Q_X^{111}$, $Q_Y^{111}$ and $Q_Z^{111}$. Qualitatively, NW4 shows very minor tilting, no intensity decay and no lattice expansion.

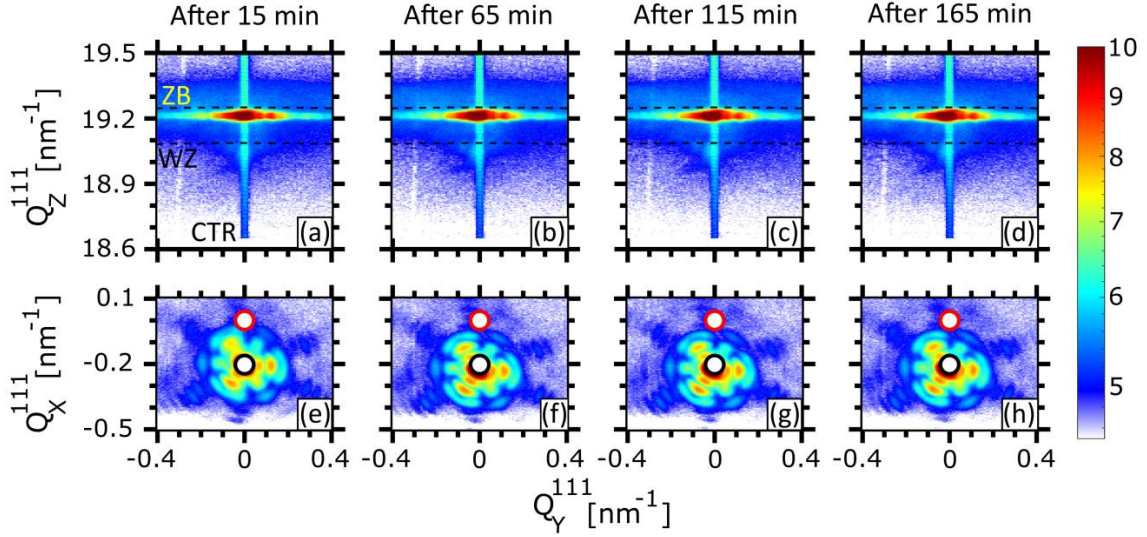

*Figure 5: (a)–(d) 2D projections of the 111 Bragg reflection of NW4, exposed under He atmosphere, in the ($Q_Z^{111}$, $Q_Y^{111}$) reciprocal space plane. The two dashed curves in $Q_Z^{111}$ represent the ZB (top) and WZ (bottom) Debye-Scherrer rings. (e)–(h) 2D projections of the 111 Bragg reflection in the ($Q_Y^{111}$, $Q_X^{111}$) reciprocal space plane. The red and black circles represent the positions of the Si CTR and the initial NW tilt, respectively. The time at which each RSM acquisition was started is mentioned at the top.*

In order to quantify the NW tilt, the RSMs in the ($Q_Y^{111}$, $Q_X^{111}$) plane were integrated along $Q_Y^{111}$ and $Q_X^{111}$ each at a time and the resulting integrated line-scans were fitted by multi-Gaussian functions. In Figs. 6(a),6(b), the angular tilts of the NW along $Q_Y^{111}$ and $Q_X^{111}$ are denoted by $α_y$ and $α_x$, respectively, tracing their variation as a function of exposure time and thus of the absorbed X-ray dose. Details on the calculation of the absorbed dose are given in the supporting information (part 3). At an exposure time of 48 min, it can be clearly seen that for NW2, P and T tilt in opposite directions indicating a small bending. The heavily illuminated part P tilts by 0.1° in $α_y$ and 0.4° in $α_x$ with respect to the initial position, whereas the less illuminated part T tilts by -0.2° in $α_y$ and $α_x$ by the end of exposure. The intensity decay of the 111 Bragg reflection is plotted in Fig. 6(c) showing an exponential decay. This curve has been calculated by integrating the intensity distribution of the 3D Bragg reflection along all three reciprocal space vectors, $Q_Z^{111}$, $Q_Y^{111}$ and $Q_X^{111}$, while excluding the Si CTR. Finally, from the variation of the Bragg reflections in $Q_Z^{111}$, we were able to calculate the variation in the axial *c* lattice parameter, $Δc$ as shown in Fig. 6(d). The unstrained axial lattice parameter of ZB GaAs (a = 5.65325 Å) was considered as a reference value ($\Delta c = 0$). The WZ polytype is formed in the upper part of the NW (see Fig. S5a in the supporting information) when the Ga-droplet is consumed at the end of the core growth (Rieger *et al.* 2013, Lähnemann *et al.* 2019). The WZ Bragg reflection undergoes a rapid thermal lattice expansion before disappearing after 92 min of exposure. Considering the thermal expansion coefficient of bulk GaAs, $6.4 \times 10^{-6}$ K$^{-1}$ (Straumanis *et al.* 1965, Pierron *et al.* 1966), the measured peak shift is associated with sample heating by about $\Delta T \approx 450$ K (see Fig. S5(b) in the supporting information). Based on SEM images, the evolution of this peak can be caused by tilting away from the Bragg condition followed by a melting of this section. The main ZB Bragg reflection, visible at the beginning of exposure, splits into 2 sub-peaks after 48 min and then into 3 sub-peaks after 92 min. Since the three sub-peaks originate from the main ZB reflection, they are referred to as ZB1, ZB2 and ZB3 in Fig. 6(d). After 92 min of exposure time, a thermal expansion gradient of about 0.5% is calculated comparing ZB3 and ZB1. As it can be seen in Fig. 6(d), ZB3 undergoes the highest lattice expansion compared to the position of the original ZB reflection at the beginning of exposure. This reflects the impact of the peak and tail exposures of the primary beam on the ZB polytype as a function of its spatial position along the NW growth axis. Therefore, ZB3 is assumed to be the region located directly below the WZ segment and beneath it is ZB2 and then ZB1 as sketched in Fig. S5(a) in part 4 of the supporting information. NW4, represented by black data points in Fig. 6, shows only negligible variation in $\alpha_y$ and $\alpha_x$, no decay in the integrated intensity of its measured 111 Bragg reflection and no axial lattice variation.

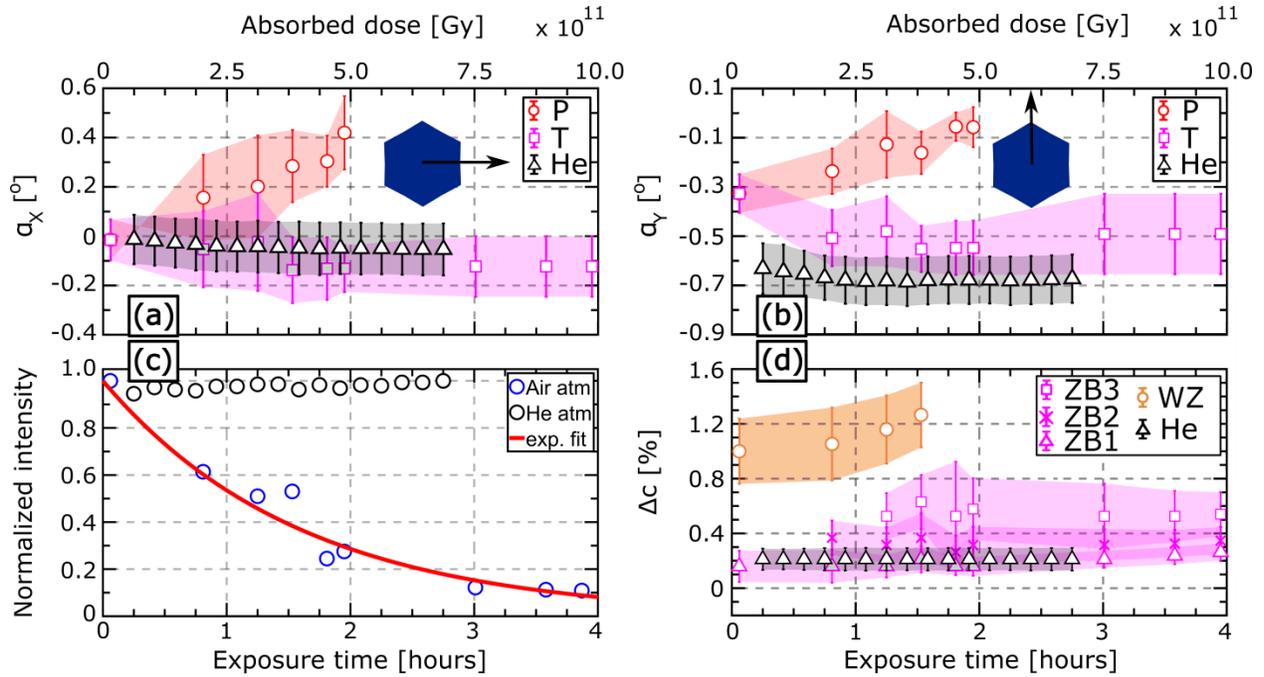

*Figure 6: (a) and (b) tilt of NW2 (air) and NW4 (He) along $Q_X^{111}$ and $Q_Y^{111}$, respectively. The blue hexagon represents the NW cross-section and the black arrow represents the tilt direction. Red data points correspond to section P, whereas the pink ones correspond to section T of NW2. The black points correspond to NW4, exposed for 3 hours under He atmosphere. (c) Intensity decay of the integrated 111 Bragg reflection for the two NWs. The red curve is an exponential fit. Error bars are within the size of the dots. (d) Variation in the c lattice parameter. The orange points correspond to the WZ reflection. The three ZB sub-peaks, colored in pink, were named ZB1–ZB3 as indicated in Fig. 4d.*

Overall, a total of 12 NWs under air atmosphere and 6 NWs in He have been systematically exposed for different time intervals. The described results are representative for all the measured NWs. The speed at which the degradation in air proceeded varied between NWs, as differences in the alignment of the beam with respect to the NW lead to variations in the deposited dose rate. The diameter of all NWs exposed in air increased by 50–150 nm (30-100% of initial diameter), while the CL was fully quenched after less than two hours. Besides NW2, melting was observed in one more NW exposed for 3 hours. For the NWs exposed under He atmosphere, barely any morphological changes or variation of the axial lattice parameters was consistently observed, while the degradation of the CL was significantly slowed down.

**Discussion**

Our experiments have revealed significant radiation damage incurred for single GaAs/(In,Ga)As/GaAs NWs exposed in the as grown geometry by an X-ray nanobeam with a photon flux of about $10^{10}$ s$^{-1}$, provided by a 3$^{rd}$ generation synchrotron radiation facility under ambient conditions. The damage observed in air atmosphere, which is characterized by oxidation, morphological deformation and optical degradation, was significantly reduced when flushing the NWs by a continuous flow of He gas. Thus, the damage in air can be associated with the impact of ozone generated by the highly intense X-ray beam and the subsequent formation of oxide layers on the NW side planes (Flinn et al. 1990, Lu et al. 1993, Hollinger et al. 1994). Lu *et al.* (1993) have concluded that ozone-induced oxidation of GaAs (100) results in a stack of binary oxide layers, namely $Ga_2O_3$, $As_2O_3$, $As_2O_5$. In contrast, Hollinger *et al.* (1994) have shown that an amorphous, single-phase, non-stoichiometric ternary oxide is formed. This transformation of GaAs to an amorphous oxide leads to structural damage, which is in turn responsible for the loss in CL signal. Furthermore, it may inhibit the sufficient dissipation of the heat created by the X-ray beam into the surrounding and result in a heating of the NWs.

The loss of CL emission from the embedded QW after X-ray exposure is a direct consequence of the oxidation process. The CL intensity is highly sensitive to the presence of non-radiative defects. The interface of the growing oxide shell to the NW core is likely to induce strong non-radiative recombination (Küpers *et al.* 2019). However, the oxidation process may even introduce non-radiative centers deeper in the NW, for example by in-diffusion of O. Such defects could account for the locally reduced emission in the central segment of NW1 and the quenching of the luminescence in parts of NW4, where the morphological changes are more subtle. As the oxide shell grows at the cost of the outer GaAs barrier layer, the interface approaches the QW, which will increase the loss of carriers. Even before the 30 nm thick shell is completely transformed and the oxidation of the QW starts, a complete loss of the CL signal can be expected. While the heating of the NW during the prolonged X-ray exposure could at some point lead to an interdiffusion of Ga and In atoms that degrades the previously well-defined QW shell, the emission energy of such a wider (In,Ga)As shell with reduced In content would shift to higher energies, which is not observed experimentally. Therefore, we conclude that the oxidation is likely to be the main driving force for the quenching of the CL emission.

Concerning the melting of NW segments exposed for an extended period in air, the peak shifts in our nXRD experiments indicate a heating of up to 450 K. Wallander *et al.* (2017) investigated the effect of X-ray-induced NW heating using time-resolved finite-element modeling. They studied the time-dependent response of single and repeated pulsed X-rays on an InP NW, either dispersed on a conductive $Si_3N_4$ membrane or in the as grown geometry. Thus, they could demonstrate that a thermal equilibrium in the NW during X-ray exposure is reached within tens of nanoseconds. The heat transport towards the substrate was identified as the main channel of heat dissipation for NWs dispersed horizontally on the substrate, as the temperature generated within the NW in the as grown geometry was significantly larger. For the as grown NWs, the authors concluded that heat transfer to the surrounding atmosphere through convection becomes the dominant cooling channel as the contact area to the substrate is small.

Additionally taking into account the oxide layer, this finding can also explain our results. The strong structural damage up to the melting of the top section observed under air atmosphere for NW2 clearly reflects that the temperature within NWs that have been exposed to X-rays under ambient conditions must be much higher compared to the NWs exposed in He atmosphere. While thermal equilibrium is reached on a nanosecond timescale, the structural damage occurs on the same slow timescale (hours) as the buildup of the oxide shell. It is thus reasonable to assume that the oxidation affects the heat dissipation and thereby leads to an increase in temperature. The formation of an oxide shell with lower heat conductivity would further reduce the contact area with the substrate. Even more importantly, the heat transfer to the surrounding atmosphere may be reduced by such a shell, which acts as a thermal resistor for heat transport.

Neglecting the continuous axial heat flow, the NW temperature created by the reduced heat transfer through the barrier can be estimated considering the 2D solution of the heat transfer equation, namely the temperature gradient through a heat resisting barrier. It is given by (Meschede *et al.* 2004):

$$T_{NW} = T_0 + \frac{Pd}{A\lambda} \qquad (1)$$

where $P$ is the heating power from the X-ray beam, $A$ is the illuminated surface area (the cross-section irradiated by the X-ray beam), $\lambda$ is the heat conductivity and $d$ is the thickness of the barrier constituted by the oxide shell formed around the NW surface. The thermal conductivity coefficient of GaAs $\lambda^{GaAs}$ is about 50 W/mK (Maycock *et al.* 1967, Sze *et al.* 1981). $\lambda^{\beta-Ga_2O_3}$ of crystalline $Ga_2O_3$ is found to range between 9.5 and 22.5 W/mK depending on orientation (Guo *et al.* 2015, Handwerg *et al.* 2016), while for amorphous oxides and amorphous semiconductors $\lambda$ is expected to be at least one order of magnitude smaller (Yoshikawa *et al.* 2013, Wingert *et al.* 2016). Since $T_{NW}$ is linear in $d$, but inversely proportional to $\lambda$, the increase in temperature can be explained by the increasing thermal resistance of the growing oxide layer. Based on equation 1, a numerical estimate provides a figure for $T_{NW} - T_0$ on the order of 100 K for a conservative estimate of $\lambda$. but a higher value if a smaller value for $\lambda$ is assumed (see section 4 of the supporting information). On the other hand, considering the massive electronic excitation by the x-ray beam and subsequent electron-electron and electron-phonon interaction where a major part of the energy is transferred to the lattice, non-thermal melting may additionally contribute to the increase of the lattice temperature. These estimates support our conclusion that the observed shift of the Bragg reflections is by various sources of heating for samples covered by an oxide layer. In contrast, NWs exposed under He atmosphere show a reduced oxide deposition and subsequently a higher heat transfer to the surrounding He atmosphere, which keeps the temperature of the NW at a lower level. In addition, the heat dissipation from the NW side facets to the gas atmosphere through convection is more effective under He compared to air because of the higher thermal conductivity of He (0.1513 W/mK) compared to air (0.024 W/mK) (Yang *et al.* 2010).

## Summary:

We have demonstrated that continuous illumination of compound semiconductor NWs under air atmosphere by an X-ray beam focused to a sub-micron spot size at a fixed beam position can induce profound changes in the structure and morphology of the NWs with severe impact on the emission properties of embedded core-shell QWs as evidenced by CL measurements. As demonstrated, the exposition of group-III arsenide NWs under ambient conditions for one hour leads to a local degradation of the CL emission and to an about 30% increase of the NW diameter. Major structural changes, including tilting, lattice expansion and an intensity reduction of the Bragg reflection, already set in at this stage. NWs exposed for 2 hours show no more CL emission. When the exposure time exceeds 3 hours, a melting of parts of the NWs is possible. The damage to the NWs can be largely mitigated by measurements under He atmosphere. For exposures up to 3 hours, neither a variation in the axial lattice spacing, nor an intensity decay of the measured Bragg reflection are observed. The CL is quenched at the exposed area, but not necessarily along the whole length of the NW. Therefore, and with the help of EDX measurements, we attribute the structural changes to oxidation of GaAs under the influence of ozone created by the X-ray beam under ambient conditions. This oxidation limits the heat transfer from the NW side facets to the surrounding atmosphere and leads to significant sample heating and thus lattice expansion and eventually a melting of parts of the NW. The reduced CL emission is attributed to nonradiative recombination at the interface between GaAs and amorphous oxides, probably combined with the introduction of nonradiative centers deeper in the NW.

The presented results have a significant impact on experiments to be performed in future nano-beam stations at $3^{rd}$ and $4^{th}$ generation synchrotron facilities. In general, it would be desirable to avoid beam damage by performing the experiment under inert gas atmosphere as shown in this report. This approach may be complemented by an improved contact between the NW and the substrate, a continuous sample cooling or an attenuation of the beam. With the emergence of Artificial Intelligence into scientific applications and analysis, raw 2D intensity detector frames of these diffraction patterns, with proper metadata context, could be used as training data for Artificial Intelligence recognition of damage onset using machine learning algorithms, forming a new scope of automation for future scattering experiments.

## Acknowledgements:

This work was supported by DFG under grant Pi218/38. We acknowledge DESY (Hamburg, Germany), a member of the Helmholtz Association HGF, for the provision of experimental facilities. Parts of this research were carried out at the PETRA III beamline P08.

# Supporting Information for Beam damage of single semiconductor nanowires during X-ray nano beam diffraction experiments

Ali AlHassan[(1),*], Jonas Lähnemann[(2),*], Arman Davtyan[(1)], Mahmoud Al-Humaidi[(1)], Jesús Herranz[(2)], Danial Bahrami[(1)], Taseer Anjum[(1)], Florian Bertram[(3)], Arka Bikash Dey[(3)], Lutz Geelhaar[(2)], Ullrich Pietsch[(1)]

[(1)]Naturwissenschaftlich-Technische Fakultät der Universität Siegen, 57068 Siegen, Germany
[(2)]Paul-Drude-Institut für Festkörperelektronik, Leibniz-Institut im Forschungsverbund Berlin e.V., Hausvogteiplatz 5-7, 10117 Berlin, Germany
[(3)]DESY Photon Science, Notkestr. 85, 22607 Hamburg, Germany

(*) Both authors, Ali Al Hassan and Jonas Lähnemann, contributed equally to this work.

## Part 1: He implementation

In order to replicate the nano X-ray diffraction (nXRD) experiment under He atmosphere, a cylinder made from kapton tape was used to shield the sample from the ambient conditions inside the experimental hutch. A picture of this configuration is shown in Fig. S1. A constant He flux enters through the blue tube and leaves through a hole in the side of the cylinder so that a constant pressure is maintained inside the cylinder.

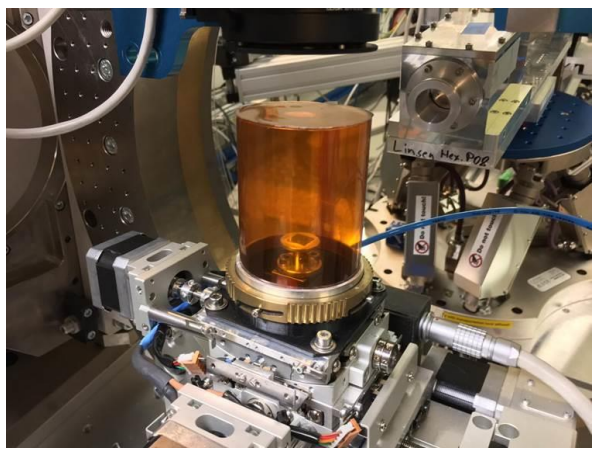

*Figure S1: Kapton tape cylinder implemented to shield the sample from air atmosphere. The He gas enters through the blue tube. A hole was drilled at the side wall of the Kapton tape to release the He pressure.*

## Part 2: Reciprocal space maps for NW1

NW1 was exposed to X-rays for 1 hour in air. The reciprocal space maps (RSMs) recorded during this time are shown in Fig. S2 and show a variation of the Bragg peak along $Q_Y^{111}$ only. Similar as during the first hour of exposure for NW2, the main peak splits into 2 sub-peaks P (peak) and T (tail). This behavior indicates that NW1 does not undergo any thermal expansion and moreover tilts only in one direction perpendicular to the [111] growth direction. Quantitatively, the NW is tilted by 0.4° to 0.6° from the initial orientation (Fig. S3).

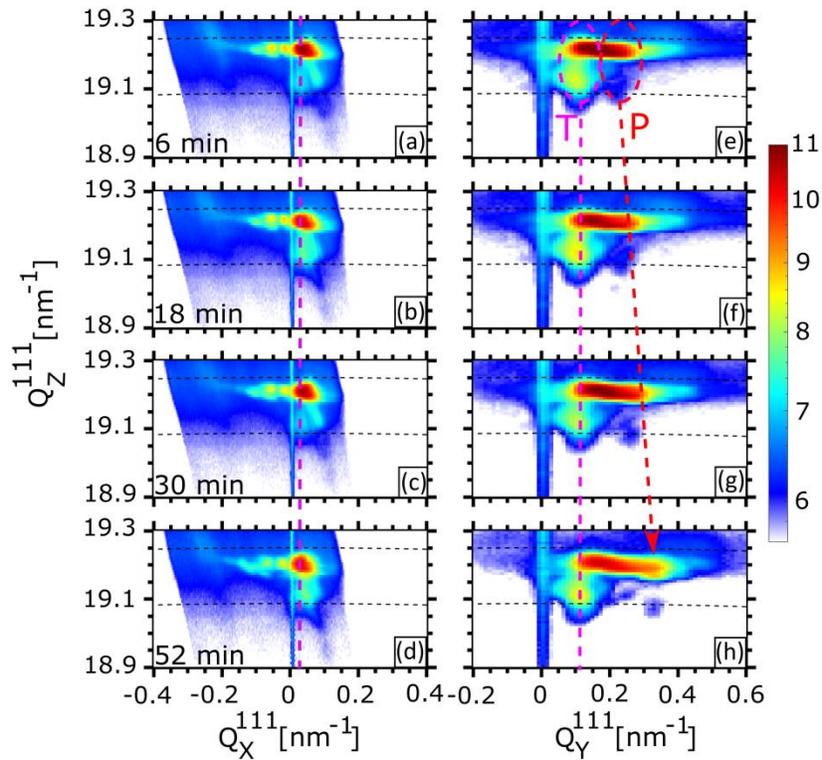

*Figure S2: (a-d) RSMs in the ($Q_Z^{111}$,$Q_X^{111}$) plane and (e-h) in the ($Q_Z^{111}$,$Q_Y^{111}$) plane during exposure. The time at which the RSM acquisition was started is mentioned at the bottom left corner of each sub-plot. The pink dashed lines indicate the constant positions of T in $Q_X^{111}$ (a-d) and $Q_Y^{111}$ (e-h), whereas the red dashed arrow indicates the variation of P in $Q_Y^{111}$ (e-h).*

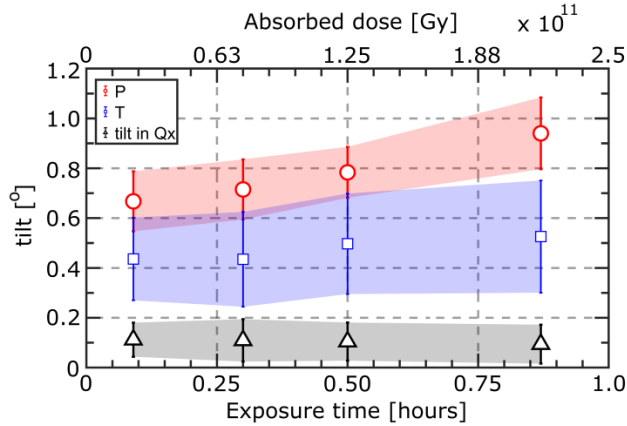

*Figure S3: Tilt calculation of P and T ($Q_Y^{111}$ direction), as well as in $Q_X^{111}$ direction, calculated from the RSMs in Figure S2.*

## Part 3: Calculation of the absorbed dose

As shown in the main text, the structural changes within single nanowires (NWs) were monitored and displayed as a function of exposure time (hours) and absorbed dose rate (Gy). In order to calculate the absorbed dose, first the cross-section of the experimentally used Gaussian nano-focused beam was reconstructed knowing that the total photon flux is $10^{10}$ s$^{-1}$ and that the vertical and horizontal full width at half maxima of the beam are 600 nm and 1800 nm, respectively. Second, a single NW with nominal height of 2 μm and diameter of 150 nm was created and convoluted with the Gaussian beam as depicted in Fig. S4(a). The simulation considers a possible misalignment of the NW with respect to the beam during the experiment.

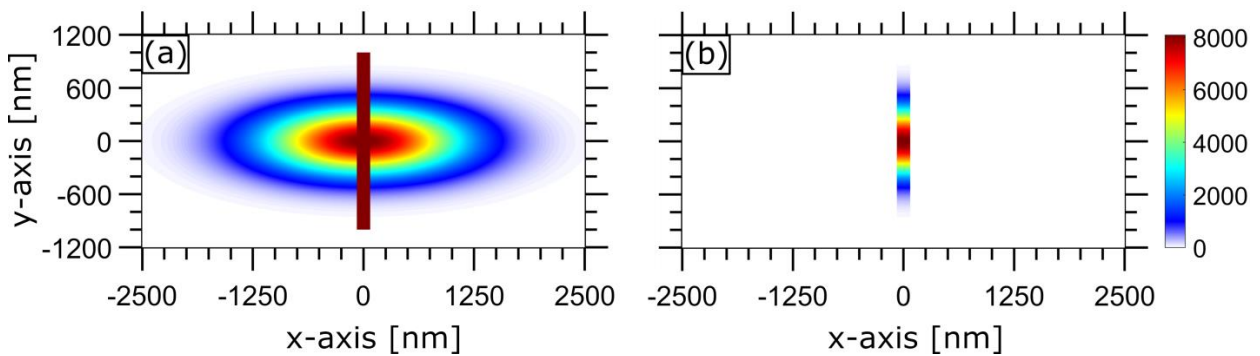

*Figure S4: (a) The reconstructed Gaussian beam aligned to the center of a NW with nominal height of 2000 nm and diameter of 150 nm. (b) The number of photons illuminating the NW if the Gaussian beam hits the center of the NW.*

Integrating the signal in Fig. S4(b), the photon flux that illuminates the NW ($F^{ill}$) is calculated to be $5.94 \times 10^8 \, s^{-1}$. The dose is measured in Gy (J/kg), requiring the NW mass in kg and the deposited energy in J as input.

The total flux that hits the NW, expressed in units of J, depends on the number of photons with photon energy $E_{ph} = 9$ keV and is,

$$F^{tot} = F^{ill} \times E_{ph} \times e = F^{ill} \times 9 \times 10^3 \times 1.6 \times 10^{-19} \, J = 8.55 \times 10^{-7} \, J/s \qquad (1)$$

Considering the nominal NW thickness of 150 nm and a height of 600 nm, given by the vertical full width at half maximum of the beam, the illuminated NW volume ($V^{ill}$) is calculated to be $1.17 \times 10^7 \, nm^3$. The mass, $M$, of the NW is calculated using $V^{ill}$ and the density of GaAs ($\rho^{GaAs} = 5318 \, kg/m^3$) as

$$M = V^{ill} \times \rho^{GaAs} = 6.22 \times 10^{-17} \, kg. \qquad (2)$$

The dose rate per second ($D_0$) that hits the NW is defined as,

$$D_0 = F^{tot}/M = 1.37 \times 10^{10} \, Gy/s. \qquad (3)$$

The dose rate absorbed by the NW requires the NW thickness $d$, which is chosen to be the nominal NW diameter of 150 nm, and the linear absorption coefficient $\mu$, which is $3.37 \times 10^{-5} \, nm^{-1}$. Using these parameters, the absorbed dose rate per second $D_A$ and the transmitted dose rate per second $D_T$ can be calculated using equations 4 and 5,

$$D_A = D_0 \times (1 - e^{-\mu d}) = 6.94 \times 10^7 \, Gy/s. \qquad (4)$$

$$D_T = D_0 \times e^{-\mu d} = 1.37 \times 10^{10} \, Gy/s. \qquad (5)$$

## Part 4: Estimate of sample heating due to growth of an oxide layer

Using equation (1) from the main manuscript, one may estimate the difference in temperature between a heated NW ($T_{NW}$) and the surrounding ($T_0$) through a heat resistor layer of thickness $d$. Using $P = F_{tot}$; an oxide layer of $d = 100$ nm and an approximated heat transfer coefficient $\lambda = 1$ W/mK (Yoshikawa *et al.* 2013, Wingert *et al.* 2016) for the amorphous oxide shell, as well as the illuminated area equal to the total spot size (150 nm × 600 nm), we estimate

$$T_{NW} - T_0 = \frac{Pd}{A\lambda} = \frac{(9\times10^{-7}W)(10^{-7}m)}{1\frac{W}{mK}(9\times10^{-14})m^2} \approx 1K$$

This number is too small to explain a significant NW heating. However, considering that the major damage is found at the center of the Gaussian beam, $A$ is reduced by a factor 100 to an area of about 900 nm² and the temperature change increases to 100 K. If the true photon flux is higher than estimated or the heat transfer coefficient is even lower, this number increases further. Thus, we can infer that NW melting can become possible for oxide thicknesses on the order of several ten nanometers.

Fig. S5(a) shows a sketch of the NW core-shell configuration and Fig. S5(b) the temperature increase for NW2. The temperature change presented in Fig. S5(b) has been calculated from the variation in the axial lattice variation (Δc) of WZ, ZB1, ZB2 and ZB3 in Fig. 6(d), where the reference position corresponding to ΔT = 0 is that of the main ZB and WZ Bragg peaks in Fig. 4(a). As demonstrated, the temperature can reach values above the congruent decomposition temperature of GaAs of approximately 680 ºC (Cheyn *et al.* 2015) and close to the melting point of GaAs of approximately 1240 ºC (Dhanaraj *et al.* 2010) with an error of about ± 230 ºC calculated by error propagation taking into account the uncertainties of the axial lattice parameter.

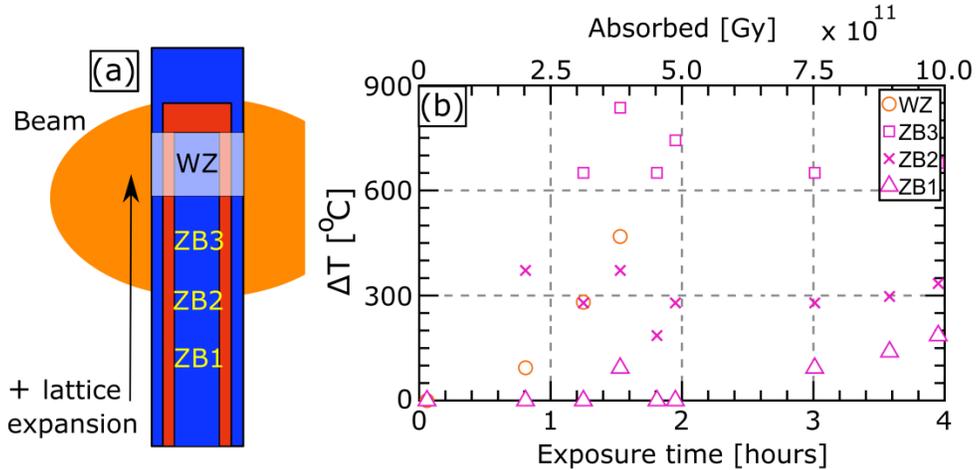

*Figure S5: (a) Side view sketch showing the core-shell-shell configuration of a single NW. These NWs typically grow in the ZB crystal phase, whereas the WZ polytype is present in the upper part of the NW, just below the top section formed by axial elongation during shell growth. (b) Change in lattice temperature with exposure time (or absorbed radiation dose) extracted from the positions of the WZ and ZB Bragg reflections.*